\documentclass{aastex63}

\usepackage{graphicx}
\usepackage{subfigure}

\usepackage{mathtools}

\usepackage{cleveref}

\usepackage{booktabs}
\usepackage{multirow}

\usepackage{xspace}

\newcommand{\dd}{\mathbf{d}}
\newcommand{\Ep}{$E_\text{p}$\xspace}
\newcommand{\Mbulk}{$M_\text{bulk}$\xspace}
\newcommand{\Be}{$B''_\text{e}$\xspace}
\newcommand{\gm}{$\gamma_\text{e,m}$\xspace}

\begin{document}

\title{Gamma-ray Burst Prompt Emission Spectrum and $E_\text{p}$ Evolution Patterns in the ICMART Model}

\author[0000-0002-2694-3379]{Xueying Shao}
\affiliation{Department of Astronomy, Beijing Normal University, Beijing 100875, China}

\author[0000-0002-3100-6558]{He Gao}
\affiliation{Department of Astronomy, Beijing Normal University, Beijing 100875, China}

\correspondingauthor{He Gao}
\email{gaohe@bnu.edu.cn}

\begin{abstract}
    In this paper, we simulate the gamma-ray bursts (GRBs) prompt emission light curve, spectrum and \Ep evolution patterns within the framework of the Internal-Collision-induced MAgnetic Reconnection and Turbulence (ICMART) model.
    We show that this model can produce a Band shape spectrum, whose parameters (\Ep, $\alpha$, $\beta$) could distribute in the typical distribution from GRB observations, as long as the magnetic field and the electron acceleration process in the emission region are under appropriate conditions.
    On the other hand, we show that for one ICMART event, \Ep evolution is always a hard-to-soft pattern.
    However, a GRB light curve is usually composed of multiple ICMART events that are fundamentally driven by the erratic GRB central engine activity.
    In this case, we find that if one individual broad pulse in the GRB light curve is composed of multiple ICMART events, the overall \Ep evolution could be disguised as the intense-tracking pattern.
    Therefore, mixed \Ep evolution patterns can coexist in the same burst, with a variety of combined patterns.
    Our results support the ICMART model to be a competitive model to explain the main properties of GRB prompt emission.
    The possible challenges faced by the ICMART model are also discussed in details.
\end{abstract}

% \maketitle

\section{Introduction}
\label{sec:introduction}

Gamma-ray bursts (GRBs) are the most luminous explosions in the universe.
Their bursty emission in the hard-X-ray/soft-$\gamma$-ray band is usually called the ``prompt emission" \citep{Zhang2018book}.
The temporal structure of the prompt emission exhibits diverse morphologies \citep{Fishman1995}, which can vary from a single smooth pulse to extremely complex light curves with many pulses overlapping within a short duration.
\citet{Gao2012} proposed that the prompt emission light curves are typically the superposition of an underlying slow component and a more rapid fast component,
where the fast component tends to be more significant in high energies and becomes less significant at lower frequencies \citep{Vetere2006}.

The photon number spectrum of prompt emission, both for time-resolved spectrum and time-integrated spectrum, can usually be fitted with a broken power law known as the Band function \citep{Band1993}
\begin{equation}
    N(E) = \begin{cases}
        A(\frac{E}{100\text{keV}})^\alpha \text{e}^{-E/E_0} &, E<(\alpha-\beta)E_0\\
        A[\frac{(\alpha-\beta)E_0}{100\text{keV}}]^{\alpha-\beta} \text{e}^{\beta-\alpha} (\frac{E}{100\text{keV}})^\beta &, E \ge (\alpha-\beta)E_0
    \end{cases}
    \label{eq:band function}
\end{equation}
where $A$ is the normalization factor, $E_0$ is the break energy in the spectrum, $\alpha$ and $\beta$ are the low-energy and high-energy photon spectral indices with $\alpha$ ranging at $(-2, 0)$ and $\beta$ ranging at $(-4, -1)$ \citep{Preece2000}.
The peak energy of the $E^2N(E)$ spectrum ($E_p = (2+\alpha) E_0$) distributes within several orders of magnitude but clusters around $200-300 \text{keV}$ \citep{Preece2000,Goldstein2013}.
For bright bursts, two types of evolution patterns between \Ep and flux are usually observed, i.e. the hard-to-soft pattern \citep{Norris1986} where \Ep decreases throughout the pulse and the intensity-tracking pattern \citep{Kargatis1994,Bhat1994,Golenetskii1983} where \Ep tracks the radiation intensity.

After decades of investigations, the origin of GRB prompt emission is still under debated. The main obstacle in front of theorists is the composition of GRB outflows.
For matter-dominated scenario, the most representative model is the so-called the internal shock model \citep{Rees1994}.
In this model, a relativistic unsteady outflow is generated from the central engine, which can be represented as a succession of shells ejected with random velocity, mass and width.
Mechanical collisions between the faster late shells and the slower early shells could convert a certain fraction of the relative kinetic energy into internal energy by the resulting internal shocks.
The internal energy thus generated is then radiated via synchrotron or inverse Compton scattering and giving rise to the prompt emission.
In this scenario, the non-thermal spectrum is naturally expected as long as the collision radius is beyond the photosphere radius, and the temporal variability is attributed to the erratic activity of the central engine while the angular spreading time of shells at different radius causes the variable timescales \citep{Rees1994,Kobayashi1997}.
This model is difficult to account for superposed slow and fast variability components, unless the central engine itself carries these two variability components in the time history of jet launching, which has no proper explanation in physics \citep{Hascoet2012}.
For the evolution patterns between \Ep and flux, the internal shock model is easy to explain intensity-tracking pattern, but difficult to interpret the hard to soft pattern.
In addition, the internal shock model also faces other challenges \citep[][for a review]{Zhang2011},
including the efficiency problem \citep{Panaitescu1999,Kumar1999}, fast cooling problem \citep{Ghisellini2000,Kumar2008}, the electron number excess problem \citep{Daigne1998,Shen2009}, the missing bright photosphere problem \citep{Zhang2009,Daigne2002} and so on.

In order to solve these problems, people proposed that GRB outflows should be Poynting flux dominated instead of matter-dominated.
Significant magnetic dissipation may happen to power GRB prompt emission through different mechanisms, including MHD-condition-broken scenarios \citep{Usov1994,Zhang2002}, radiation-dragged dissipation model \citep{Meszaros1997}, the slow dissipation model \citep{Thompson1994,Drenkhahn2002,Giannios2008,Beniamini2017}, current-driven instabilities \citep{Lyutikov2003}, and forced magnetic reconnection \citep{Zhang2011,McKinney2012,Lazarian2019}.
In this work, we focus on one representative model, i.e., the Internal-Collision-induced MAgnetic Reconnection and Turbulence (ICMART) model \citep{Zhang2011}.
This model invokes a central engine powered, magnetically dominated outflow with magnetization factor $\sigma > 1$.
Similar to the internal shock model, mechanical collisions between mini-shells would distort the ordered magnetic field and trigger fast magnetic seeds, which would induce relativistic magnetohydrodynamics (MHD) turbulence in the interaction regions\footnote{It is worth noticing that rather than the mechanical collisions, the kink instability may be the cause of the magnetic turbulence but the physics processes they involve are similar \citep{Lazarian2019}.}.
The turbulence further distorts the magnetic lines, resulting in a reconnection cascade and thus a significant release of the stored magnetic field energy (an ICMART event). The particles, either accelerated directly in the reconnection region or accelerated randomly in the turbulence region, would radiate synchrotron radiation photons to power the observed GRB prompt emission.

A global simulation of an ICMART event has been presented in \citet{Deng2015}, which shows significant energy dissipation when two highly magnetized blobs collide. Mini-jet-like reconnection events are seen from the simulation, even though local simulations are needed to display mini-jets in much smaller scales.
On the other hand, Monte Carlo simulations have been performed to simulate the prompt emission light curves within the framework of the ICMART model \citep{Zhang2014}.
They show that the ICMART model can produce highly variable light curves with both fast and slow components, where the fast component is caused by many local Doppler-boosted mini-jets due to turbulent magnetic reconnection and the slow component is due to the runaway growth and subsequent depletion of these mini-jets.
However, since these simulations were focusing on the temporal structure, they assumed that the radiation intensity arising from each reconnection event had the same spectral form (e.g., Band function with $\alpha=-1,~\beta=-3$ and $E_p=300\text{keV}$ in the observer frame), which made them ineffective in testing prompt emission spectrum properties.

In this work, we will further revise the previous simulation, by considering the evolution of the magnetic field with respect to radius and using a fast-cooling synchrotron spectrum to calculate the energy spectrum for each reconnection event, in order to judge whether the ICMART model could generate the Band spectrum and interpret the observed \Ep evolution patterns.

\section{Simulation Methods}
\label{sec:method}

Within the ICMART scenario, a highly magnetized central engine would eject an unsteady outflow with variable Lorentz factors and luminosities but a nearly constant degree of magnetization.
For simplicity, the outflow can be represented as a succession of discrete shells with variable Lorentz factors.
Mechanical collisions between mini-shells would distort the magnetic field configurations and induce MHD turbulence, which will further distort the magnetic field and eventually trigger an ICMART event.
Our simulation starts with tracking a shell with mass \Mbulk, Lorentz factor $\Gamma_0$ and magnetized factor $\sigma_0 = E_\text{m, 0}/E_\text{k, 0}$ ($E_\text{m, 0}$ marks the initial magnetic energy and $E_\text{k, 0}$ marks the initial kinetic energy), when the shell expands to radius $R_0$ and the reconnection cascade has been triggered.

Numerical simulations have shown that the reconnection-driven magnetized turbulence could self-generate additional reconnections \citep{Takamoto2015,Kowal2017,Takamoto2018}, inferring that the number of magnetic reconnections would increase exponentially. Here we assume that each reconnection event ejects a bipolar outflow and triggers two more reconnection events\footnote{Lacking detailed numerical simulations for a reconnection/turbulence cascade, the specific index of magnetic reconnection growth is still unknown. Here we assume each reconnection would trigger two more new reconnection events. The uncertainty brought by this assumption will not affect our analysis results about spectrum and $E_p$ evolution (because these results are already the superposition effect of sufficient magnetic reconnection events), but might affect the time when our simulated light curve would reach the peak.} \citep{Zhang2014}. Assuming the generation number of magnetic reconnections within the $1/\Gamma_0$ cone is $n_\text{g}$, the total number of magnetic reconnections $N_\text{cone}$ could be estimated as $N_\text{cone} \approx \sum_{n=1}^{n_\text{g}} 2^n$.
The duration for each reconnection in the lab frame could be estimated as
\begin{equation}
    T_\text{r, lab} \sim \Gamma_0 \frac{L'}{v'_\text{in}} = 10^4 s \left(\frac{\Gamma_0}{100}\right)\left(\frac{L'}{10^{11}\text{cm}}\right)\left(\frac{v'_\text{in}}{10^{9}\text{cm/s}}\right)^{-1},
\end{equation}
where $L'$ is the size of the magnetic reconnection and $v'_\text{in}$ is the inflow velocity of the magnetic field line.
Hereafter, parameters denoted with ($'$) are in the rest frame of the jet bulk.
Note that this duration timescale is about $T_\text{r, obs}\sim0.1\text{s}$ in the observer frame and therefore the reconnections cascade lasts for $n_\text{g}T_\text{r, obs}$ in the observer frame.

For each reconnection event, the magnetized factor of the reconnection area drops from $\sigma_0$ to $1$.
Thus the dissipated magnetic energy approximately equals to $(\sigma_0 - 1) / \sigma_0$ times the magnetic energy within the reconnection area.
We assume that half of the dissipated magnetic energy is used to boost the kinetic energy of the jet and the other half is initially distributed to electrons \citep{Drenkhahn2002a}, and then gets converted to photons through synchrotron radiation. Part of the photons that beams toward the observer is recorded as the prompt emission of GRB.

In order to simulate the observed GRB spectrum, three rest frames need to be invoked: $1)$ the rest frame of the jet bulk; $2)$ the rest frame of the mini-jet; $3)$ the rest frame of the observer (here we ignore the cosmological expansion effect).
Hereafter, parameters denoted with ($''$) are in the rest frame of the mini-jet.
The quantities within these three frames are connected through two Doppler factors, i.e.,
\begin{align}
    \mathcal{D}_1 &= [\Gamma(1-\beta_\text{bulk}\cos{\theta})]^{-1},   \\
    \mathcal{D}_2 &= [\gamma(1-\beta\cos{\phi})]^{-1},
\end{align}
where $\gamma \approx \sqrt{1+\sigma}$ is the relative Lorentz factor of mini-jet with respect to the jet bulk \citep{Zhang2014}, $\beta_{\rm bulk}$ and $\beta$ are the corresponding dimensionless velocities with respect to $\Gamma$ and $\gamma$, $\theta$ is the latitude of the mini-jet, i.e., the angle between the line of sight and the direction of the bulk at the location of the mini-jet and $\phi$ is the angle between the direction of the mini-jet and the direction of the bulk in the bulk comoving frame.
In our simulation, we trace each mini-jet to calculate its radiation intensity and photon energy distribution in the direction of the observer, and stack the simultaneously arriving photons from all mini-jets to obtain the overall light curve and the corresponding time-resolved spectrum.

In the rest frame of the mini-jet, synchrotron radiation power at frequency $\nu ''$ is given by \citep{Rybicki1979}
\begin{equation}
\label{eq:pnup}
P''_{\nu''} = \frac{\sqrt{3} q_e^3 B''_\text{e}}{m_{\rm e} c^2}
	    \int_{\gamma_{\rm e,m}}^{\gamma_{\rm e,M}}
	    \left( \frac{dN_{\rm e}''}{d\gamma_{\rm e}} \right)
	    F\left(\frac{\nu ''}{\nu_{\rm cr}''} \right) d\gamma_{\rm e},
\end{equation}
where $q_e$ is electron charge,
$\nu_{\rm cr}'' = 3 \gamma_{\rm e}^2 q_e B''_\text{e} / (4 \pi m_{\rm e} c)$ is the
characteristic frequency of an electron with Lorentz factor $\gamma_e$,
$B''_\text{e}$ is the comoving magnetic field strength in the emission region and $F(x)=x\int_x^\infty K_{5/3}(\xi)d\xi$ with $K_{5/3}(x)$ is the modified Bessel Function of five thirds order.
Lacking full numerical simulations of magnetic turbulence and reconnection, the comoving magnetic field strength $B''_\text{e}$ in the mini-jet is rather difficult to make a direct estimation.
Here we introduce a free parameter $k$ to connect $B''_\text{e}$ with the bulk magnetic field strength as $B''_\text{e} = \sqrt{k}B'/\gamma$.
The value of $k$ is justified by limiting $E_p$ in the simulation results to be consistent with the observational data. We assume $B''_\text{e}$ decays with the radius of the bulk as
\begin{equation}
    B''_\text{e}(R) = B''_\text{e,0}(\frac{R}{R_0})^{-b},
\end{equation}
where b would be much larger than 1, due to the rapid consumption of magnetic energy in the reconnection process.
Considering the magnetic flux conservation, the bulk magnetic field strength $B'$ decreases as the jet expanding.
The radial part of the magnetic field decreases as $B'_r \propto R^{-2}$ while the transverse part decreases as $B'_t \propto R^{-1}$.
At a large radius one has a transverse-dominated magnetic field
\begin{equation}
    B'(R) = B'_0(\frac{R}{R_0})^{-1},
    \label{eq:decaying magnetic strength}
\end{equation}
where $B'_0$ is the initial magnetic strength of the bulk.

The electrons number density could be estimated with the continuity equation \citep{Uhm2014}:
\begin{equation}
    \frac{\partial N''(\gamma_e, t'')}{\partial t''} = -\frac{\partial}{\partial \gamma_e}[\dot{\gamma_e}(\gamma_e)N''(\gamma_e,t'')]+Q(\gamma_e, t''),
    \label{eq:continuity equation of injecting electrons}
\end{equation}
where $\dot{\gamma_e}(\gamma_e)$ denotes the electron cooling rate and $Q(\gamma_e, t'')$ denotes the injected source function.
Electrons would suffer both radiative and adiabatic cooling \citep{Uhm2014}, so that
\begin{equation}
    \frac{\dd}{\dd t''}(\frac{1}{\gamma_e})=\frac{\sigma_TB_\text{e}^{''2}}{6\pi m_\text{e}c}+\frac23(\frac1{\gamma_e})\frac{\dd \ln {R}}{\dd t''},
\end{equation}
where $\sigma_T$ is the Thomson scattering cross section.
The distribution of the accelerated electrons is usually assumed to be a power-law function,
\begin{equation}
    Q(\gamma_e) = Q_0(\frac{\gamma_e}{\gamma_\text{e,m}})^{-p}, \gamma_\text{e,m}<\gamma_e<\gamma_\text{e,M},
    \label{eq:electron number distribution}
\end{equation}
where $Q_0$ is a normalization factor, \gm and $\gamma_\text{e,M}$ are the minimum and maximum injected electron Lorentz factor.
$\gamma_\text{e,M}$ could be calculated as
\begin{equation}
    \gamma_\text{e,M}=\sqrt{\frac{6\pi q_e}{\sigma_TB''_\text{e}}}.
\end{equation}
One can use electron number and energy conservation law to solve $Q_0$ and \gm for a specific reconnection event, which should require
\begin{equation}
    Q_0 (t''_\text{e}-t''_\text{s})\int_{\gamma_\text{m}}^{\gamma_\text{M}}(\frac{\gamma_e}{\gamma_\text{m}})^{-p}\dd \gamma_e = f_\text{e}\int_{t'_\text{s}}^{t'_\text{e}}{L'}^2 v'_\text{in} n'_\text{e}\dd t',\\
    \label{eq:particle number conservation}
\end{equation}
and
\begin{equation}
    \frac12 \delta E'_\text{p, inj} = \int_{t''_\text{s}}^{t''_\text{e}}\int_{\gamma_\text{m}}^{\gamma_\text{M}}Q_0(\frac{\gamma_e}{\gamma_\text{m}})^{-p}\dd \gamma_e \dd t''(\gamma_e-1)m_\text{e}c^2,
    \label{eq:particle energy conservation}
\end{equation}
where $t_\text{s}$ and $t_\text{e}$ are the starting and ending time of the magnetic reconnection, $f_\text{e}$ is the fraction of accelerated electrons within the reconnection area, $n'_\text{e} \propto R^{-2}$ is the number density of electrons in the bulk frame, and $\delta E'_\text{p, inj}$ is the dissipated energy for a single magnetic reconnection, which could be estimated as
\begin{equation}
    \delta E'_\text{p, inj} =\frac{\sigma_0-1}{\sigma_0} \int_{t'_\text{s}}^{t'_\text{s}+T'_0} \frac{B'(t)^2}{8\pi}L'(t)^2v'_\text{in}\dd t',
    \label{eq:injected magnetic energy of single MR}
\end{equation}
where $t'_\text{s}$ is the beginning time.
Considering both the expansion and reconnection effects, similar to the bulk magnetic field strength, here we assume a general evolution form for $L'$,
\begin{equation}
    L'(R) = L'_0(\frac{R}{R_0})^{l},
   \label{eq:enhancing MR size}
\end{equation}
where $L'_0$ is the initial magnetic reconnection size.

As seen by the observer, each mini-jet radiation corresponds to a single pulse starting from $t'_\text{s}/\mathcal{D}_1$ to $(t'_\text{s}+T'_0)/\mathcal{D}_1$, with radiation power
\begin{equation}
    P_{\nu}= \mathcal{D}_1^3 \mathcal{D}_2^3 P''_{\nu''}.
\end{equation}
As suggested by \citet{Zhang2014}, here we use Gaussian shape to simulate each single pulse in the light curve. For a given time interval in the observer frame ($t_1$, $t_2$), the time resolved spectrum is given by
\begin{equation}
    P_{\nu}(\nu) = \sum_{N_\text{cone}}\sum_{t=t_1}^{t=t_2} P_\nu(\nu, t)
\end{equation}
For one ICMART event, the peak time of the light curve corresponds to the total duration for the cascade process, which could be estimated as
\begin{equation}
    T_p \sim \frac{R_\text{f}}{2\Gamma^2c}\sim1.5 \text{s} \left(\frac{n_g}{10}\right)\left(\frac{\Gamma_0}{200}\right)^{-1}\left(\frac{L'}{10^{11}\text{cm}}\right)\left(\frac{v'_\text{in}}{10^{9}\text{cm/s}}\right)^{-1}.
\end{equation}
We assume an abrupt cessation of the cascade process, so that the number of new mini-jets drops to 0. The light curve after $T_p$ is therefore contributed by the high-latitude
emission from other mini-jets not along the line of sight due to the “curvature effect” delay.

\section{Results}
\label{sec:result}

In order to justify whether the ICMART model could generate the Band spectrum and the observed \Ep evolution patterns, we first run a series of simulation for one ICMART event with initial setup:
$M_\text{bulk} \in \mathcal U(3.5\times 10^{-9},3.5 \times 10^{-6})M_\odot$,
$R_0 = 10^{14}\text{cm}$, $L'_0 = 1.2\times 10^{11}\text{cm}$, $v'_\text{in} = 3 \times 10^9 \text{cm/s}$, $
n_\text{g} \in \mathcal U(13, 17)$,
$\Gamma_0 \in \mathcal U(100, 300)$,
$\sigma_0 \in \mathcal U(5, 50)$,
$k\in \mathcal U(5\times 10^{-8}, 5\times 10^{-2})$,
$p \in \mathcal U(2.3, 2.8)$,
$f_e\in \mathcal U(0.01, 1)$,
$b \in \mathcal U(1, 50)$,
$l \in \mathcal U(0, 1)$,
$\theta \in \mathcal U(0, 2/\Gamma_0)$ and $\phi \in \mathcal U(0, \frac{\pi}{2})$,
where $\mathcal U$ refers to the Uniform distribution.
In \cref{fig:compare light curves of different l,fig:fitting result,fig:compare SED of different Ep,fig:compare SED of different b and p,fig:E_peak evolution of single peak in light curve,fig:different patterns of E_peak} we plot the simulation results for selected situations that are relevant for illustrating the main conclusions, which can be summarized in the following subsections.

\subsection{Light Curve}

\begin{figure}
    \centering
    \subfigure[]{
    \includegraphics[width=0.3\textwidth]{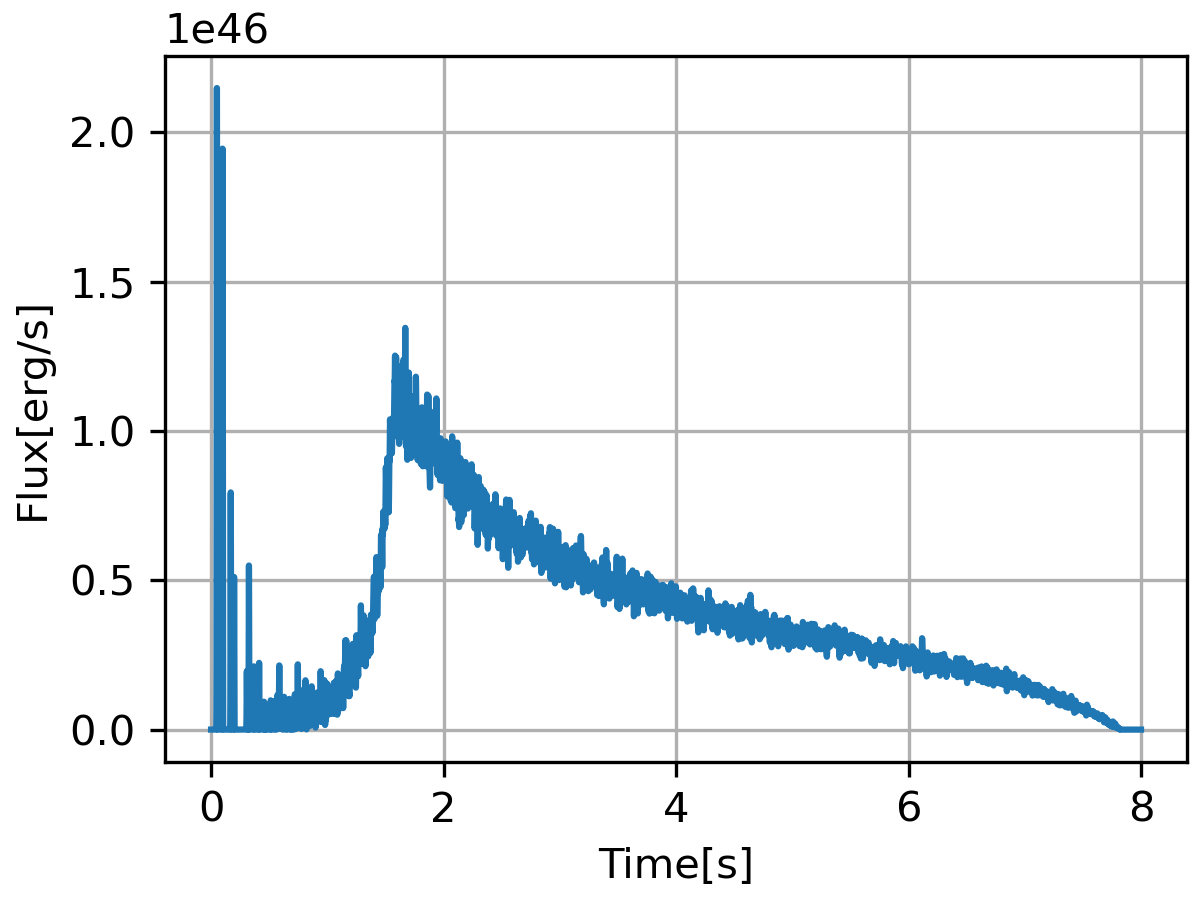}
    }
    \subfigure[]{
    \includegraphics[width=0.3\textwidth]{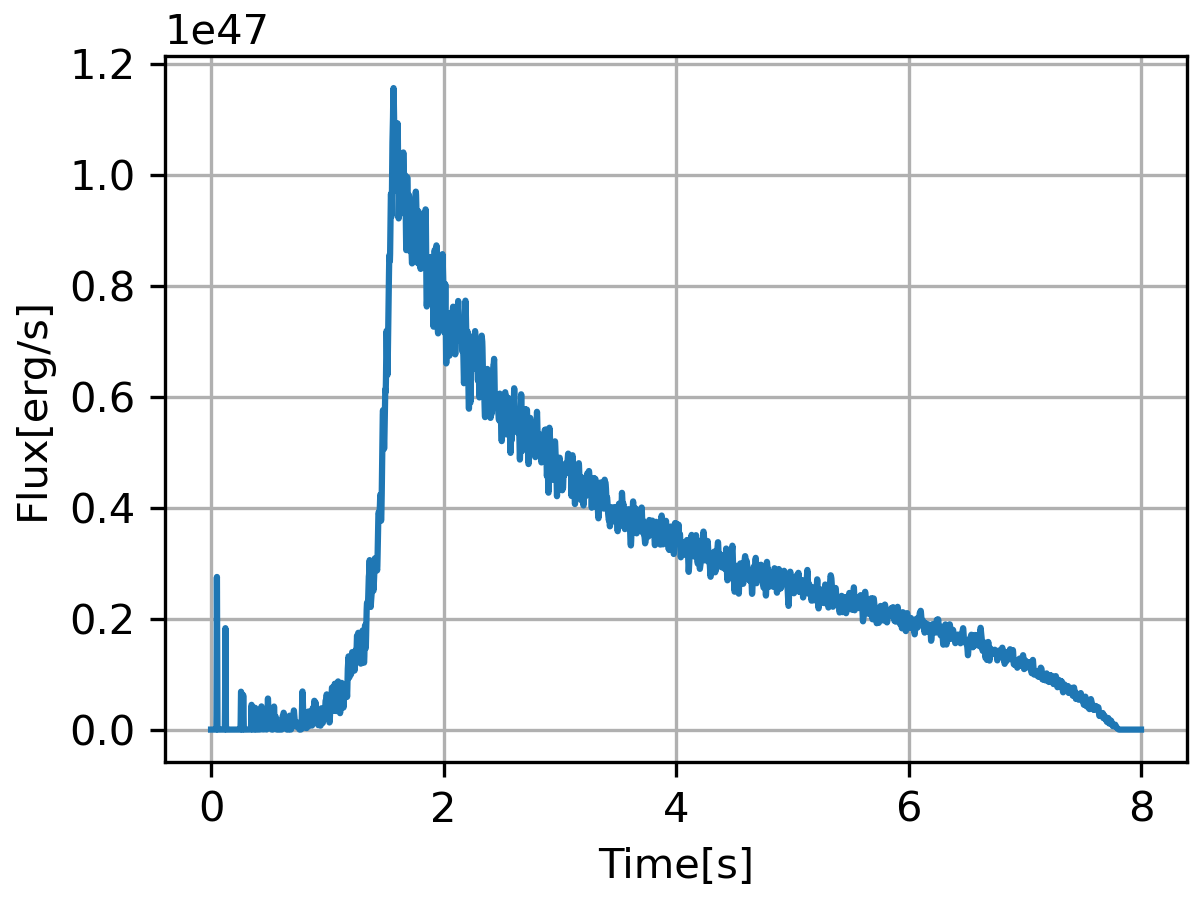}
    }
    \subfigure[]{
    \includegraphics[width=0.3\textwidth]{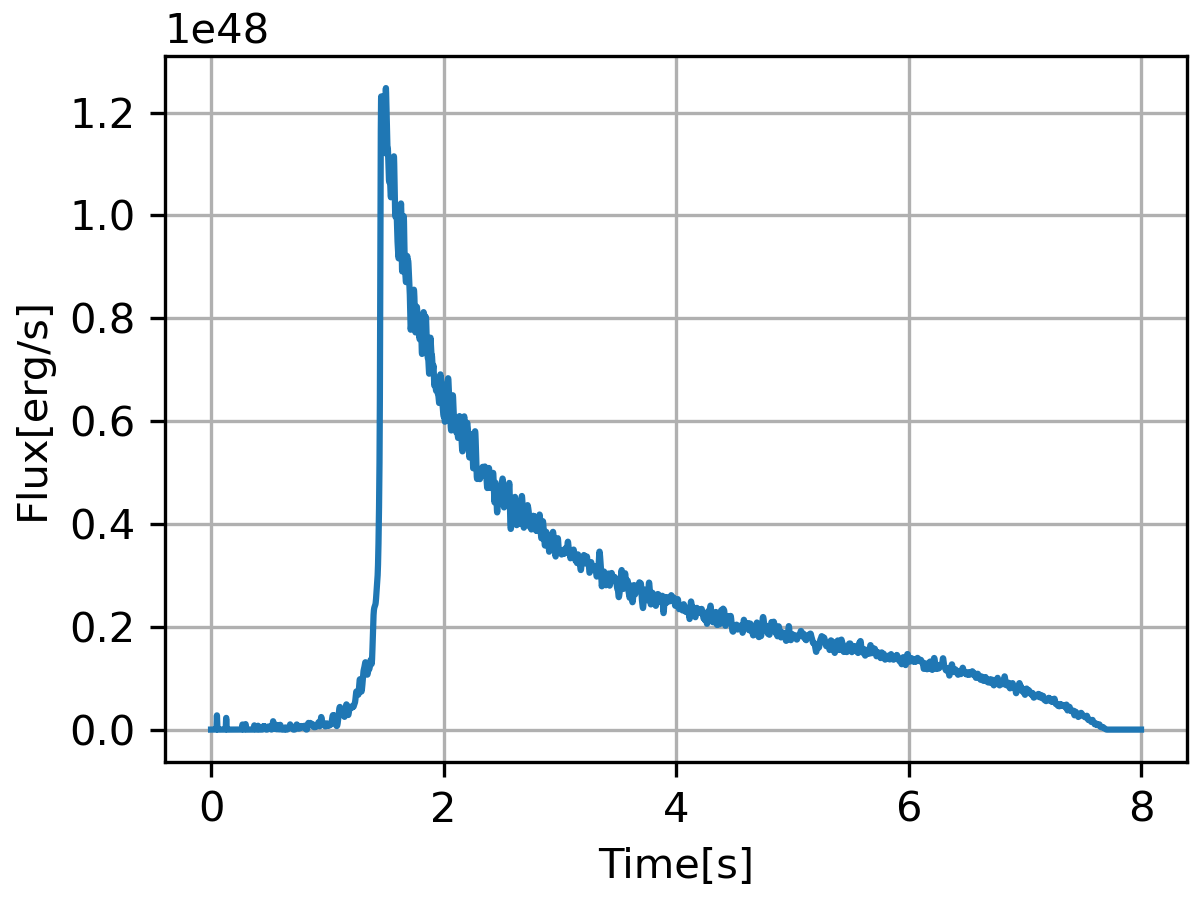}
    }
    \caption{
        Simulated light curves of one ICMART event with the following parameters: (a) $l=0.0$; (b) $l=0.2$; (c) $l=0.4$.
        }
    \label{fig:compare light curves of different l}
\end{figure}

In our simulation, each ICMART event would produce highly variable light curves, which can be decomposed as the superposition of an underlying slow component and a more rapid fast component.
The fast component is related to the individual locally Doppler-boosted mini-emitters, while the slow component is caused by the superposition of emission from all the mini-jets in the emission region.
The duration of each mini-pulse and the total duration of the light curve are essentially determined by the parameters related to the number and scale of magnetic reconnection, e.g., $L'_0$, $l$, $v'_\text{in}$ and $n_\text{g}$.

Our results are basically consistent with the previous simulation results from \citet{Zhang2014}.
It is worth pointing out that \citet{Zhang2014} did not invoke any real radiation process for each mini-jet, so they haven't discussed the evolution effect of the magnetic field and the magnetic reconnection scale. In this work, we find that if the evolution of the magnetic reconnection scale is not introduced (i.e. $l=0$), the radiation contributed by a single mini-jet in the early stage would be much larger than that in the late stage, due to the magnetic field decaying.
Consequently, some short spikes (with flux equivalent to, or even larger than the main peak) would show up in the early stage of the light curve as shown in \cref{fig:compare light curves of different l}.
This phenomenon will disappear after introducing the evolution of reconnection scale with radius (e.g. $l>0.2$).

\subsection{Spectrum}

\begin{figure}
    \centering
    \subfigure[]{
    \includegraphics[width=0.45\textwidth]{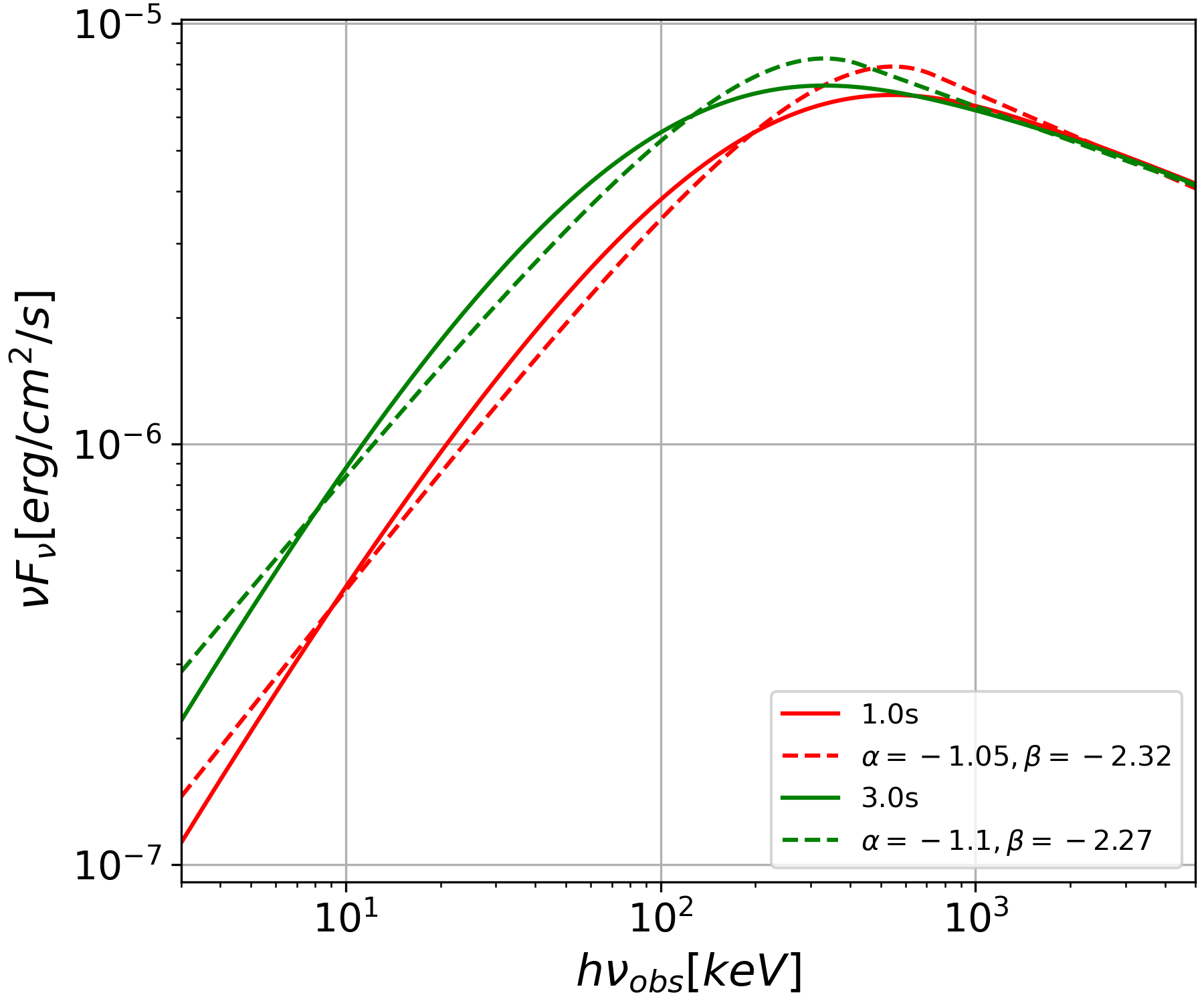}
    \label{fig:united SED}
    }
    \subfigure[]{
    \includegraphics[width=0.45\textwidth]{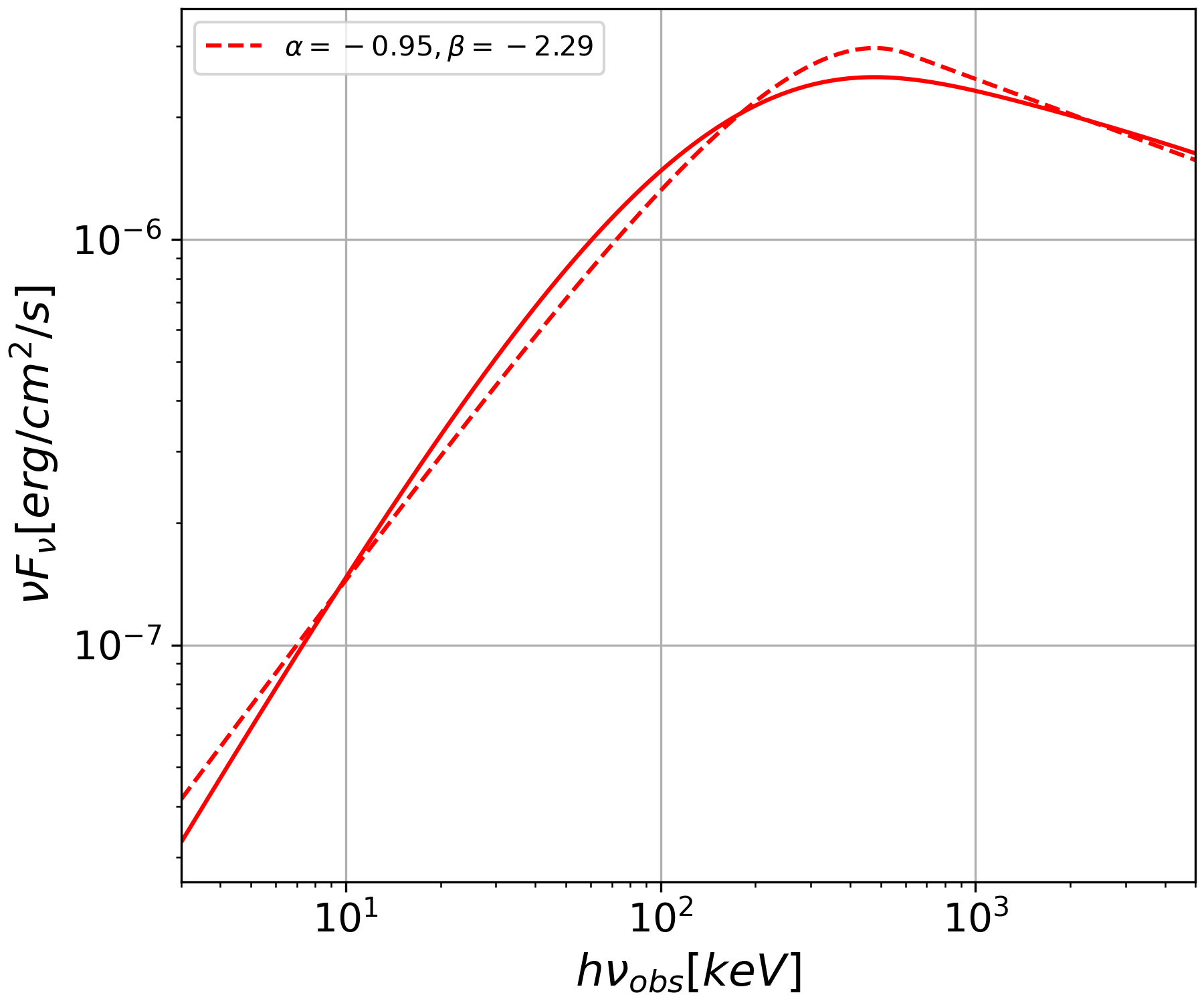}
    \label{fig:single SED}
    }
    \caption{
        (a) Simulated spectrum of one ICMART event with the following parameters: $M_\text{bulk} = 3\times 10^{-8}M_\odot$, $R_0 = 10^{14}\text{cm}$, $L'_0 = 1.2\times 10^{11}\text{cm}$, $v'_\text{in} = 3 \times 10^9 \text{cm/s}$, $n_\text{g} = 15$, $\Gamma_0 = 200$, $\sigma_0 = 8$, $k= 5\times 10^{-7}$, $p = 2.8$, $f_e=0.1$, $b =30$, $l =0.2$.
        The red and green solid lines represent the simulated spectrum at 1 s and 3 s in the observer's frame, and the dashed lines represent their relevant best fitting results with Band function.
        (b) Simulated spectrum for a single magnetic reconnection with the same parameters as the left panel but $\theta$ and $\phi$ are taken as $0$.
        The solid line represent the simulation result and the dashed line is the best fitting result with Band function.}
    \label{fig:fitting result}
\end{figure}

\begin{figure}
    \centering
    \subfigure[]{
    \includegraphics[width=0.3\textwidth]{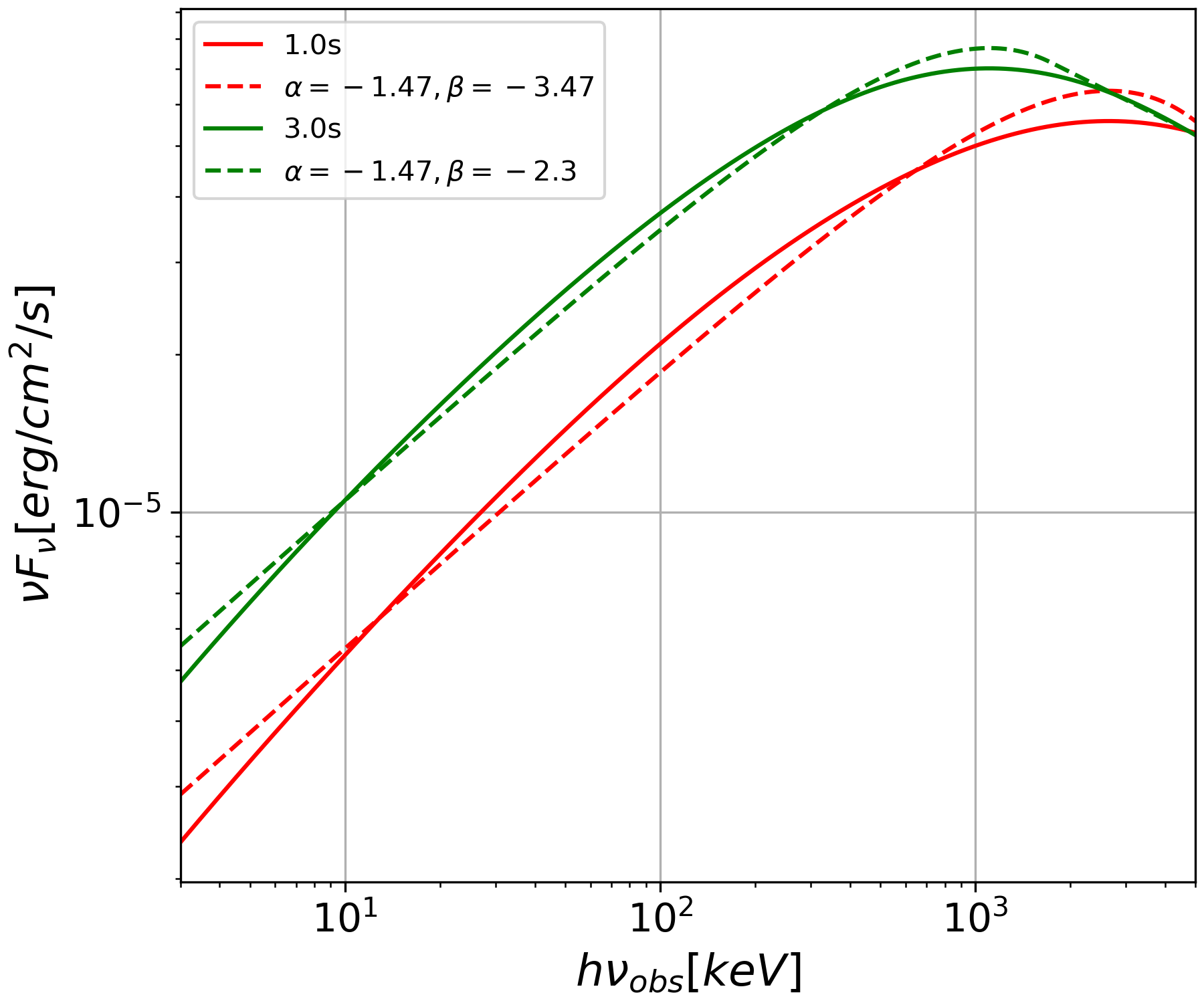}
    \label{fig:big Ep}
    }
    \subfigure[]{
    \includegraphics[width=0.3\textwidth]{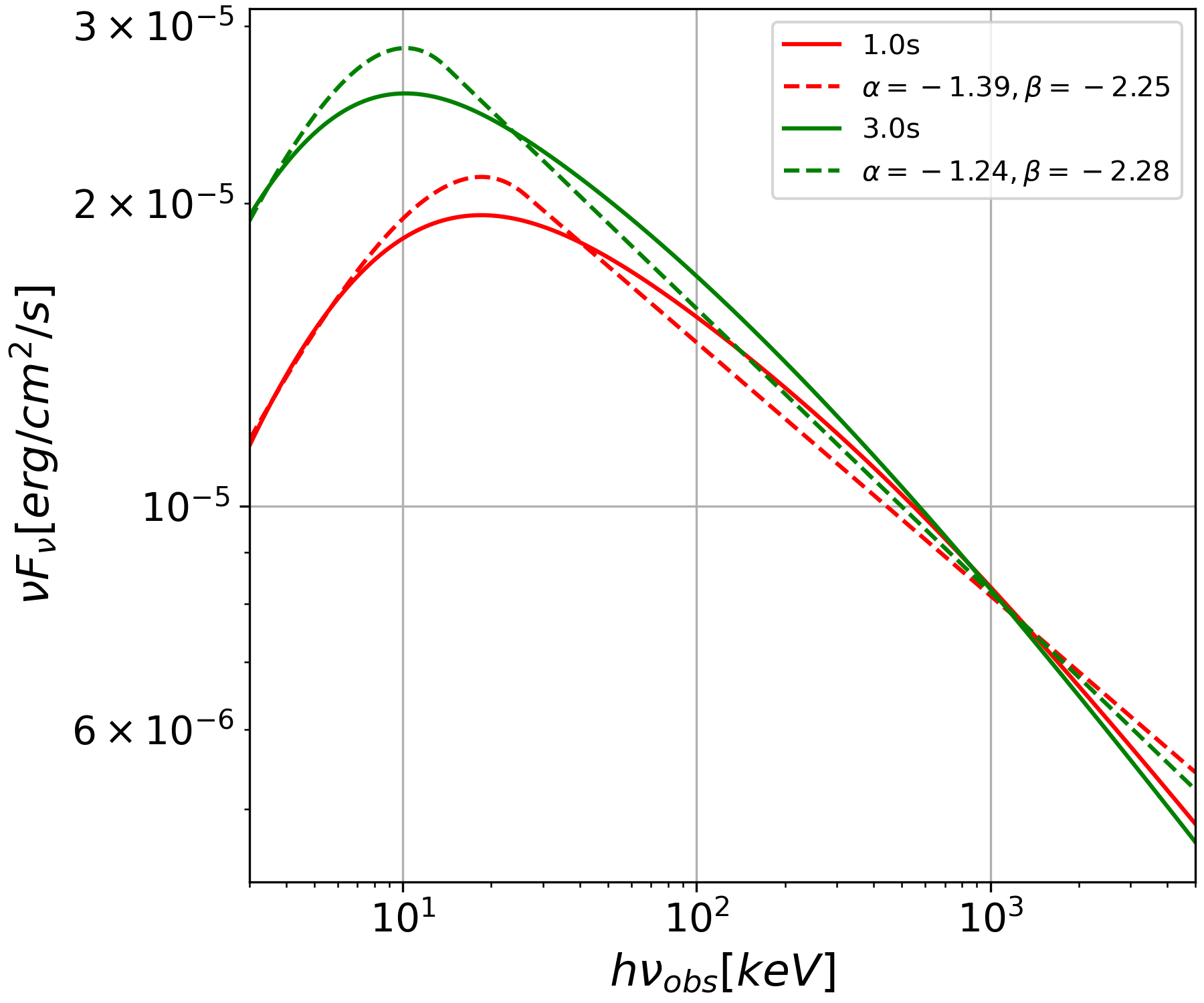}
    \label{fig:small Ep}
    }
    \subfigure[]{
    \includegraphics[width=0.3\textwidth]{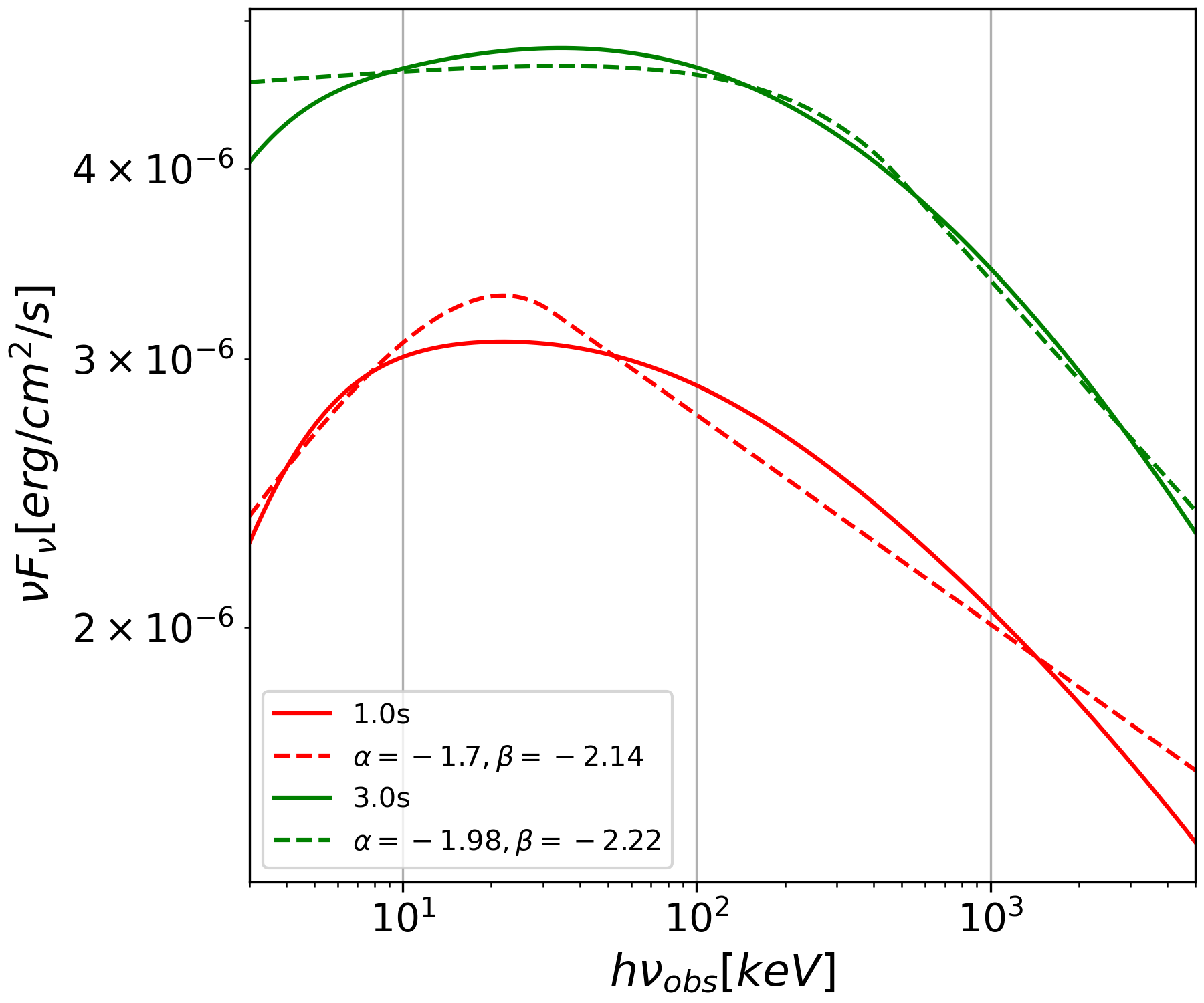}
    \label{fig:deviate Band shape}
    }
    \caption{
        Simulated spectrum of one ICMART event with the following parameters: (a) $B''_\text{e} \approx 10^3\text{G}, \gamma_\text{e,m}\approx 10^4$; (b) $B''_\text{e} \approx 500 \text{G}, \gamma_\text{e,m}\approx 10^3$; (c) $B''_\text{e} \approx 10 \text{G}, \gamma_\text{e,m}\approx 10^3$ .
        Other parameters are the same as those in \cref{fig:united SED}.
    }
    \label{fig:compare SED of different Ep}
\end{figure}

\begin{figure}
    \centering
    \subfigure[$b=1.0, p=2.8$]{
    \includegraphics[width=0.3\textwidth]{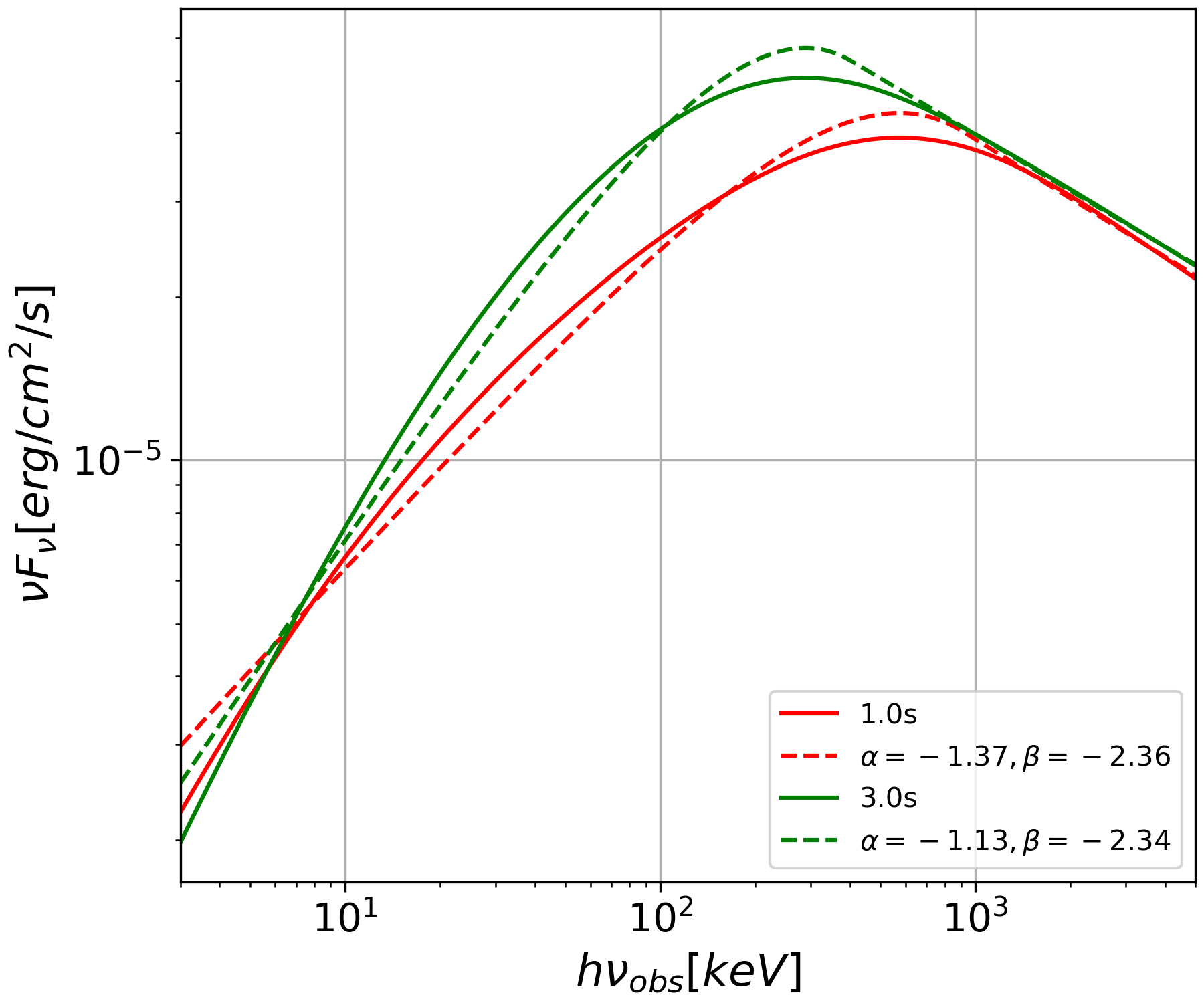}
    \label{fig:SED-1-2.8}
    }
    \subfigure[$b=30.0, p=2.8$]{
    \includegraphics[width=0.3\textwidth]{fitting_band_function_fixed-sample.png}
    \label{fig:SED-30-2.8}
    }
    \subfigure[$b=1.0, p=2.3$]{
    \includegraphics[width=0.3\textwidth]{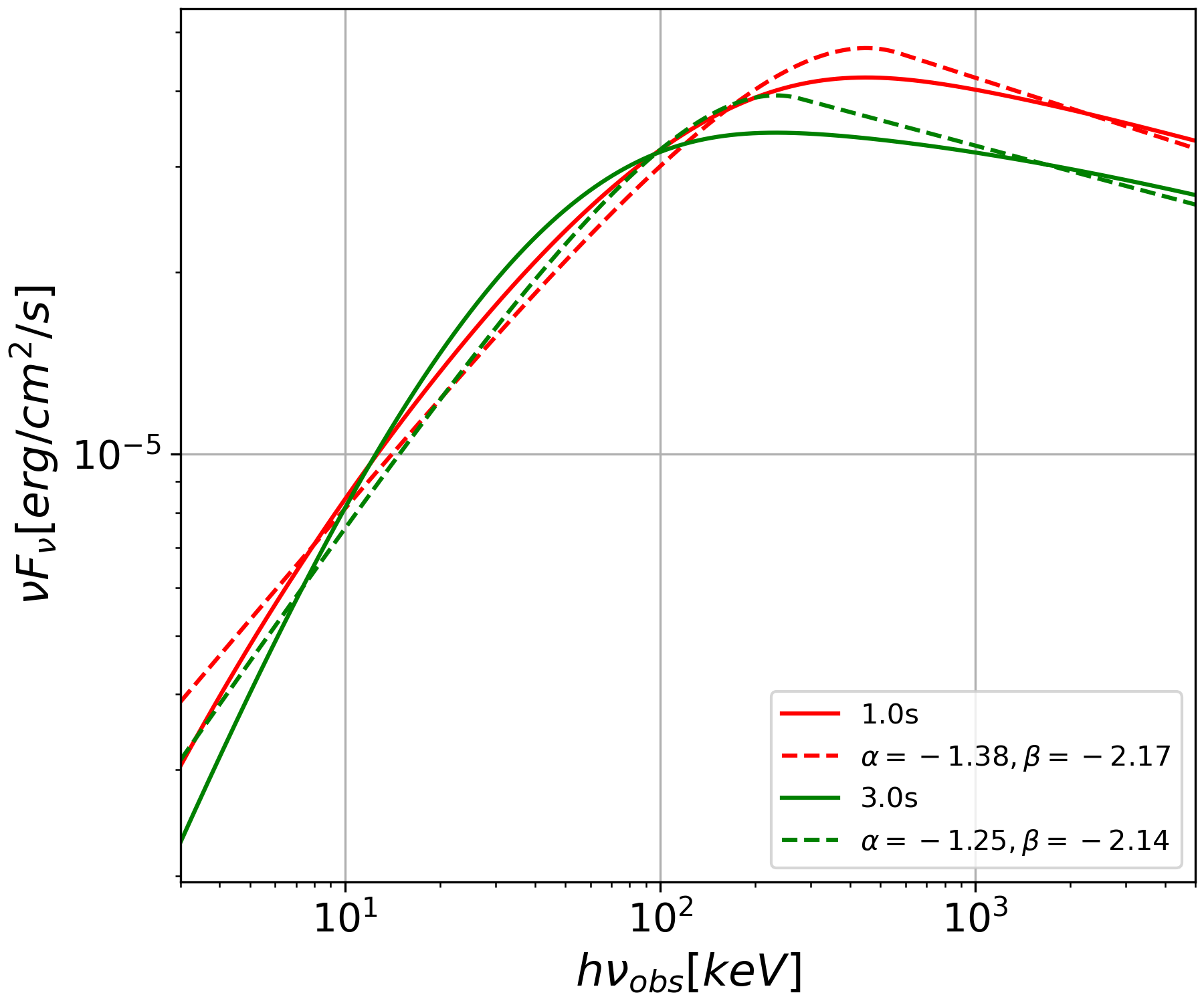}
    \label{fig:SED-1-2.3}
    }
    \caption{
        Simulated spectrum of one ICMART event with the following parameters:  (a) $b=1.0, p=2.8$; (b) $b=30.0, p=2.8$; (c) $b=1.0, p=2.3$.
        Other parameters are the same as those in \cref{fig:united SED}.
    }
    \label{fig:compare SED of different b and p}
\end{figure}

For most cases, the spectrum produced by ICMART events would behave as broken power law, which could be well fitted with the Band function.
For example, we plot the time resolved spectrum at 1 s and 3 s (in the observer frame) for one simulation
\footnote{Here we would like to note that due to the randomness of $\theta$ and $\phi$, for a given parameter combination, different simulations may give slightly different GRB light curve and spectrum, but the conclusions we show below are robust because the number of mini-jets is large enough to smooth out the main random effects.}
in \cref{fig:united SED},
in a narrower band pass from 3 keV to 3 MeV.
In the example, the best fit parameters for Band function at 1s/3s are $\alpha = -1.05/-1.1$, $\beta = -2.32/-2.27$, and $E_p=550/330$ keV, which are all typical values as suggested by the observations \citep{Preece2000}.
This result is understandable since the observed spectrum is mainly shaped by the mini-jets that beam toward the observer, while as proven by \citet{Uhm2014}, the radiation spectrum from mini-jets with $\theta=0$ and $\phi=0$ is very close to Band function (see \cref{fig:fitting result} for an example).
The simulated spectrum may be slightly broader than the spectrum of a single reconnection, which should be caused by the contributions from off-axis mini-jets.
However, the overall trend of the spectrum will not deviate too much from the Band spectrum.

Under the free combination of initial parameters, the $E_p$ value of the simulated spectrum would distribute in a very wide range, much wider than the typical distribution from GRB observations (e.g., 3 keV $\sim$ 3 MeV for current sample; \citet{Demianski2017}).
In order to make 3 keV $<E_p<$ 3 MeV, a certain degeneracy is required between the initial parameters.
For instance, since $E_p$ is essentially proportional to $\Gamma_0B''_\text{e}\gamma_\text{m}^2$ \citep{Tavani1996}, the values of $M_\text{bulk}$, $\sigma_0$, $\Gamma_0$, $R_0$, and $k$ need cooperating to make the emission region magnetic field \Be in the range of $10\sim10^4\text{G}$;
on the other hand, the values of $\sigma_0$ and $f_e$ need cooperating to make the minimum injected electron Lorentz factor \gm in the range of $10^3\sim10^5$.
Otherwise, $E_p$ would be either higher (if \Be and/or \gm being too big) or lower (if \Be and/or \gm being too small) than typical GRB $E_p$ distribution range (see \cref{fig:big Ep,fig:small Ep} for an example).
Moreover, if the magnetic field in the radiation area is too small, the electrons would be cooled inefficiently, so the spectrum may deviate from the Band shape (see \cref{fig:deviate Band shape} for an example).

For a Band-shape simulated spectrum, the higher spectral index ($\beta$) value is mainly determined by the injected electron energy distribution index ($p$), while the lower energy spectral index ($\alpha$ value) is essentially determined by the value of \Be and its evolution speed, which is reflected by the value of $b$ (see \cref{fig:compare SED of different b and p} for examples). When \Be is smaller than $1000\text{G}$ and $b$ is larger than $30$, $\alpha$ could be around $-1$ or even smaller, which is the typical value for current GRB observations. Otherwise, if \Be is too high or $b$ is too small, $\alpha$ would approach $-1.5$, entering the deep fast cooling regime for synchrotron radiation.

\subsection{\Ep Evolution}

\begin{figure}
    \centering
    \subfigure[Hard-to-Soft Pattern]{
    \includegraphics[width=0.45\textwidth]{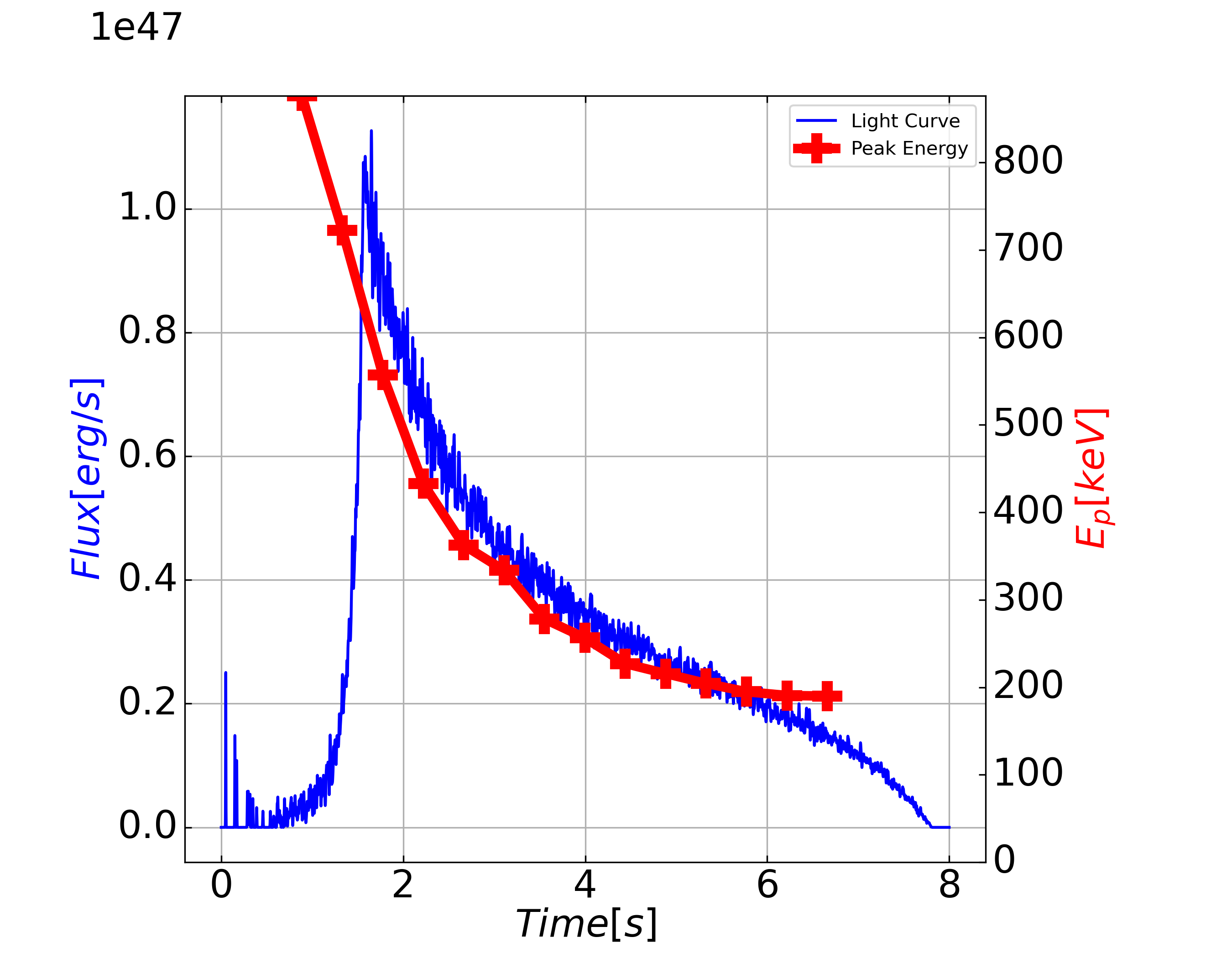}
    \label{fig:hard-to-soft E_peak evolution}
    }
    \subfigure[Tracking Pattern]{
    \includegraphics[width=0.45\textwidth]{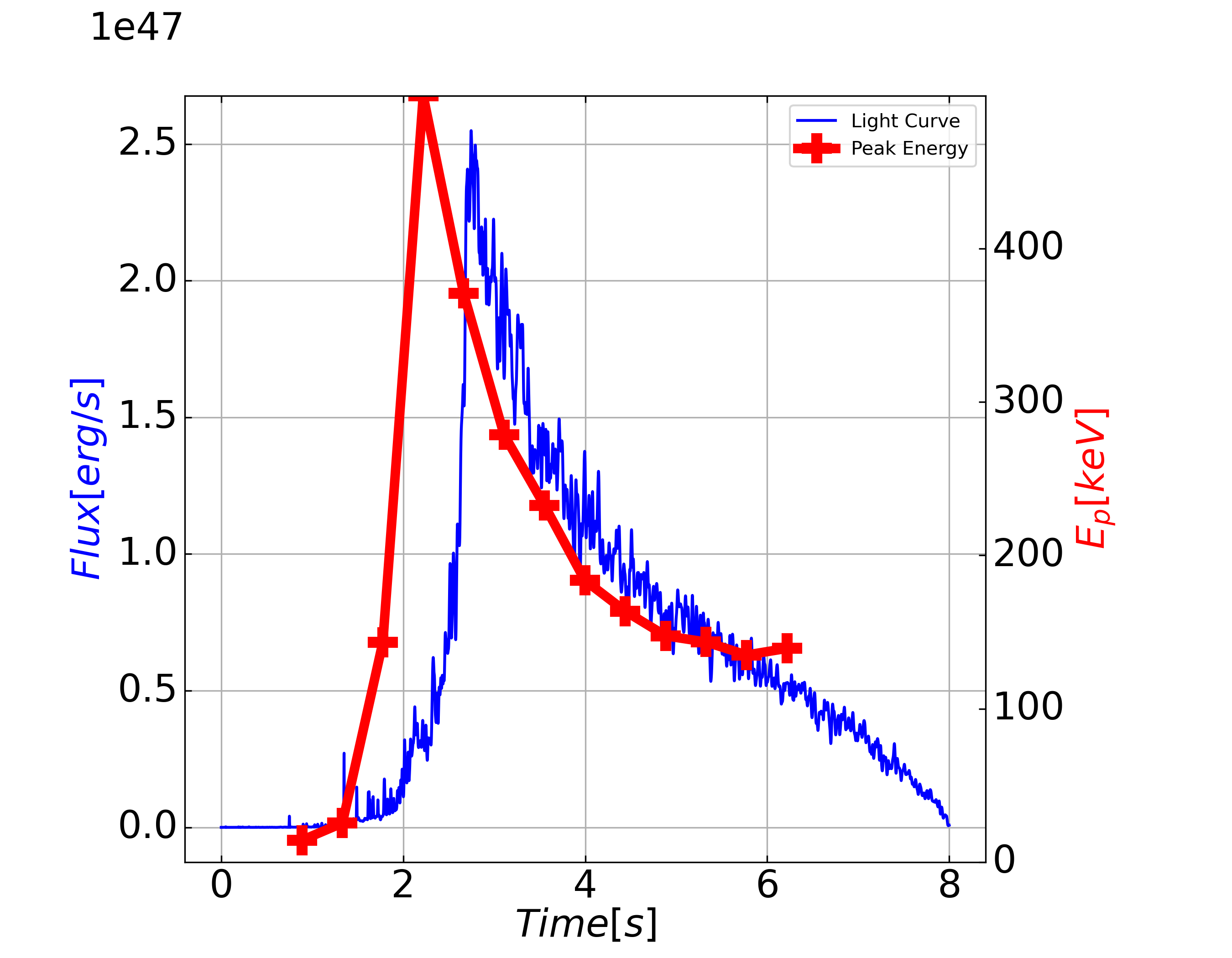}
    \label{fig:tracking E_peak evolution}
    }
    \caption{
        (a) Simulated \Ep evolution of one ICMART event, with other parameters being the same as those in \cref{fig:united SED}. (b) Simulated \Ep evolution of multiple ICMART events. The physical parameters for each ICMART event are listed in \cref{table:factors of tracking E_peak}. }
    \label{fig:E_peak evolution of single peak in light curve}
\end{figure}

\begin{deluxetable}{cccccccccc}
\tablewidth{1.0\columnwidth}
\tablehead{
\colhead{$M_\text{bulk}(M_\odot)$} & \colhead{$f_e$} & \colhead{k} & \colhead{$n_\text{g}$} & \colhead{$\sigma_0$} & \colhead{$\Gamma_0$} & \colhead{p} & \colhead{b} & \colhead{l} & \colhead{$\text{Start Time(s)}$}
}
\decimals
\startdata
$3\times 10^{-10}$ & $1.0$ & $5\times 10^{-4}$ & $15$ & $8$ & $200$ & $2.8$ & $30$ & $0.2$ & $0.0$\\
$3\times 10^{-9}$ & $0.3$ & $1.5\times 10^{-5}$ & $15$ & $8$ & $200$ & $2.8$ & $30$ & $0.2$ & $0.7$\\
$3\times 10^{-8}$ & $0.1$ & $5\times 10^{-7}$ & $15$ & $8$ & $200$ & $2.8$ & $30$ & $0.2$ & $1.3$\\
\enddata
\caption{
        Parameters for multiple ICMART events taken in the simulation of \cref{fig:tracking E_peak evolution}.
    }
    \label{table:factors of tracking E_peak}
\end{deluxetable}

\begin{figure}
    \centering
    \subfigure[]{
    \includegraphics[width=0.3\textwidth]{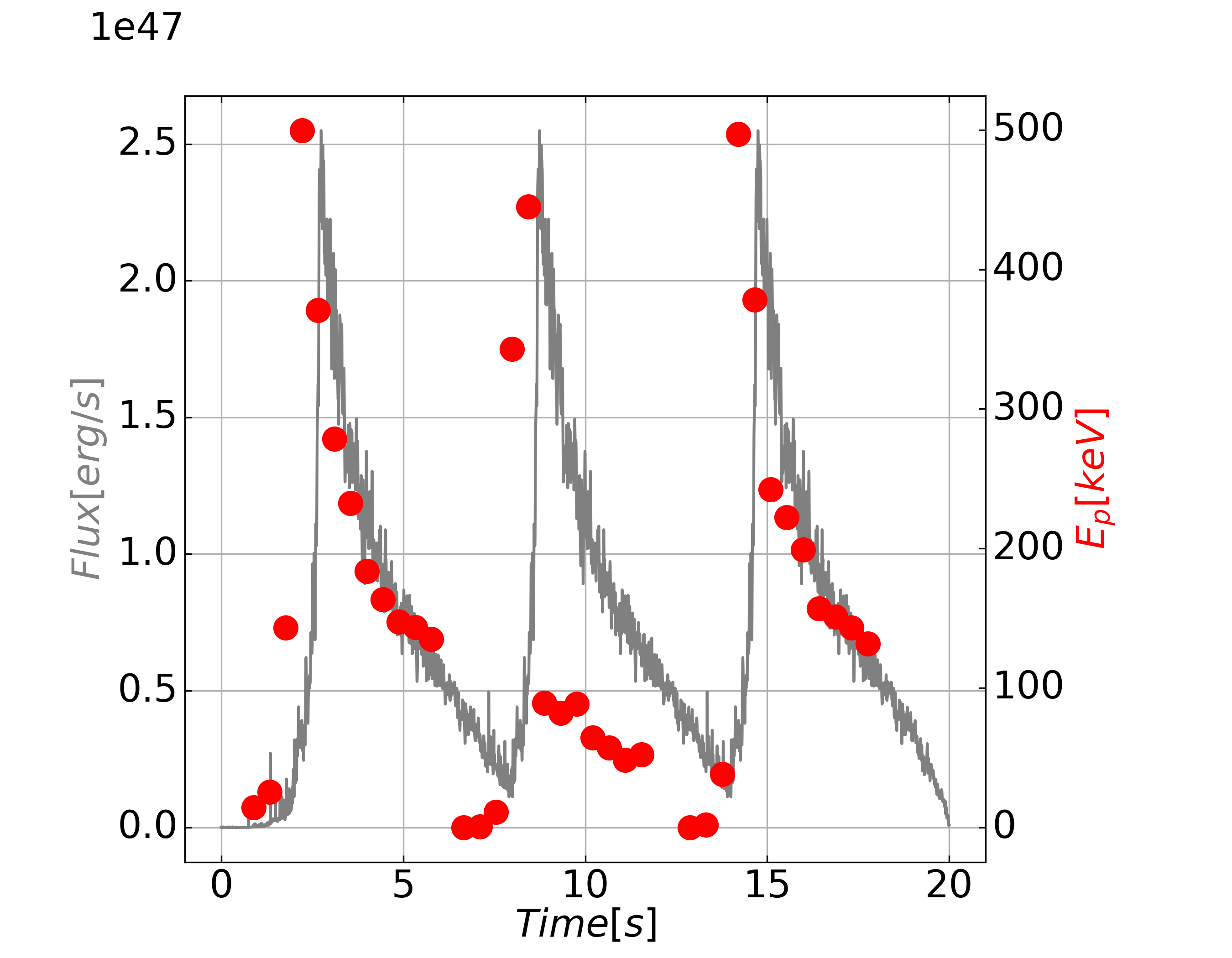}
    \label{fig:three peak}
    }
    \subfigure[]{
    \includegraphics[width=0.3\textwidth]{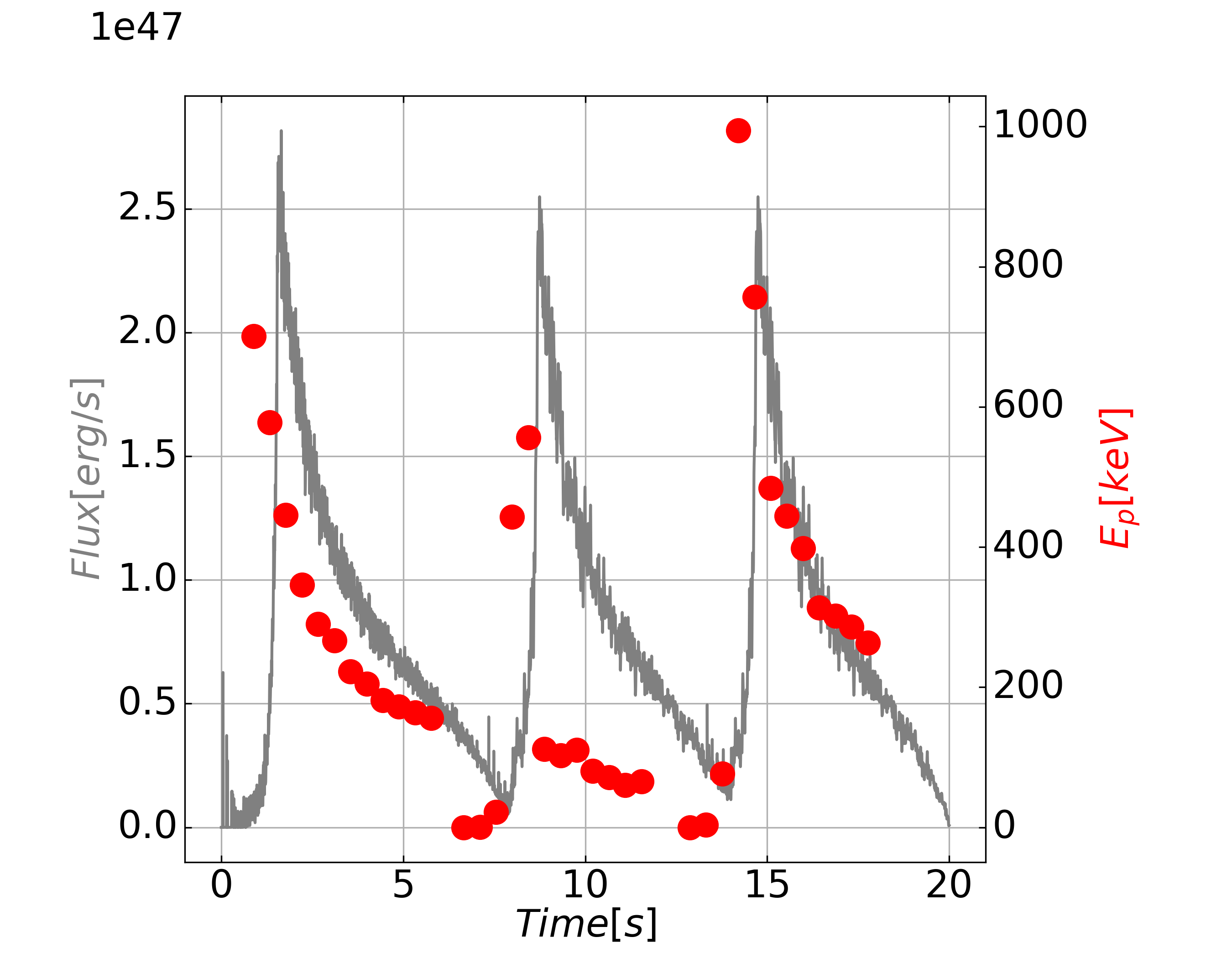}
    \label{fig:two peak}
    }
    \subfigure[]{
    \includegraphics[width=0.3\textwidth]{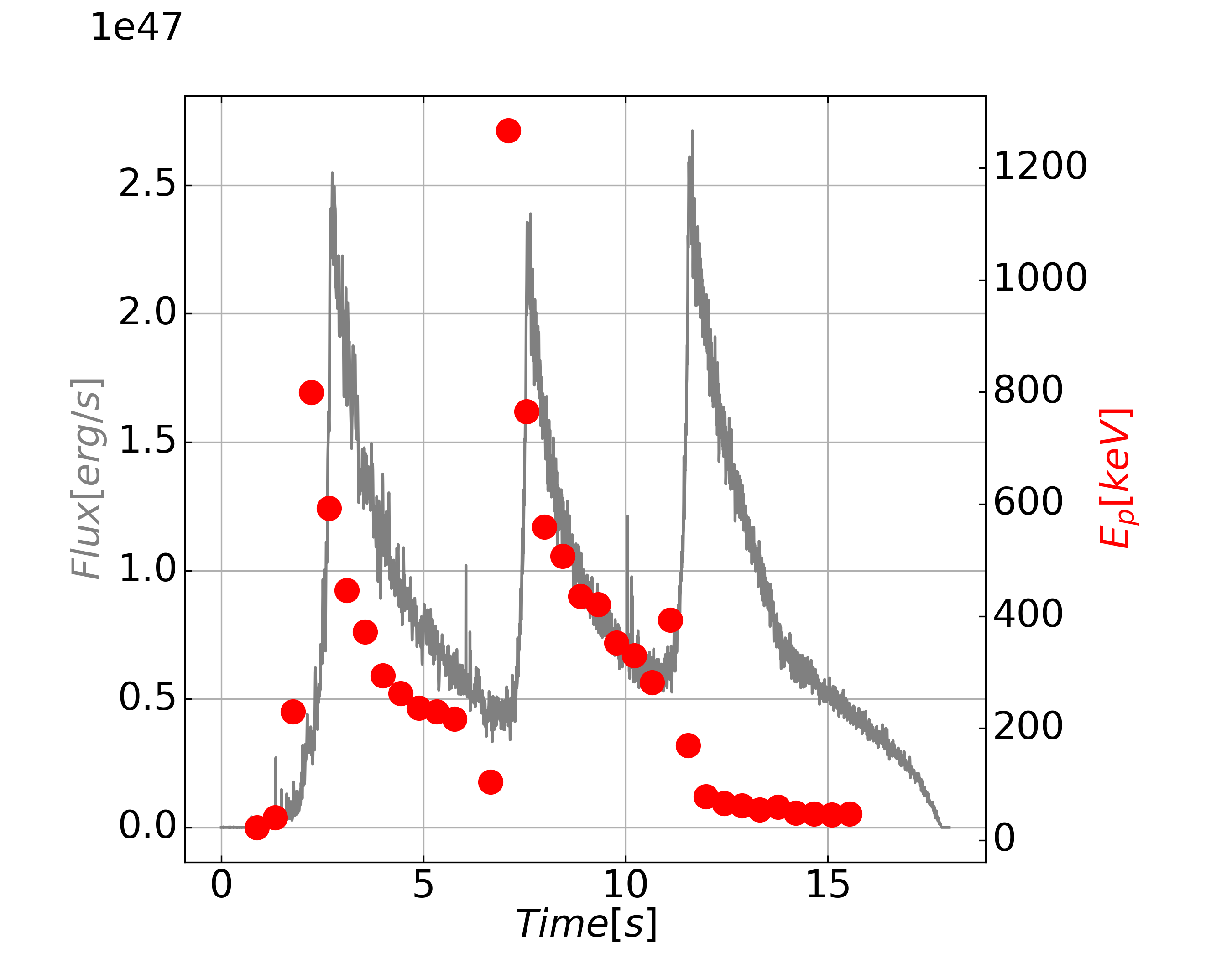}
    \label{fig:one peak}
    }
    \quad
    \subfigure[]{
    \includegraphics[width=0.3\textwidth]{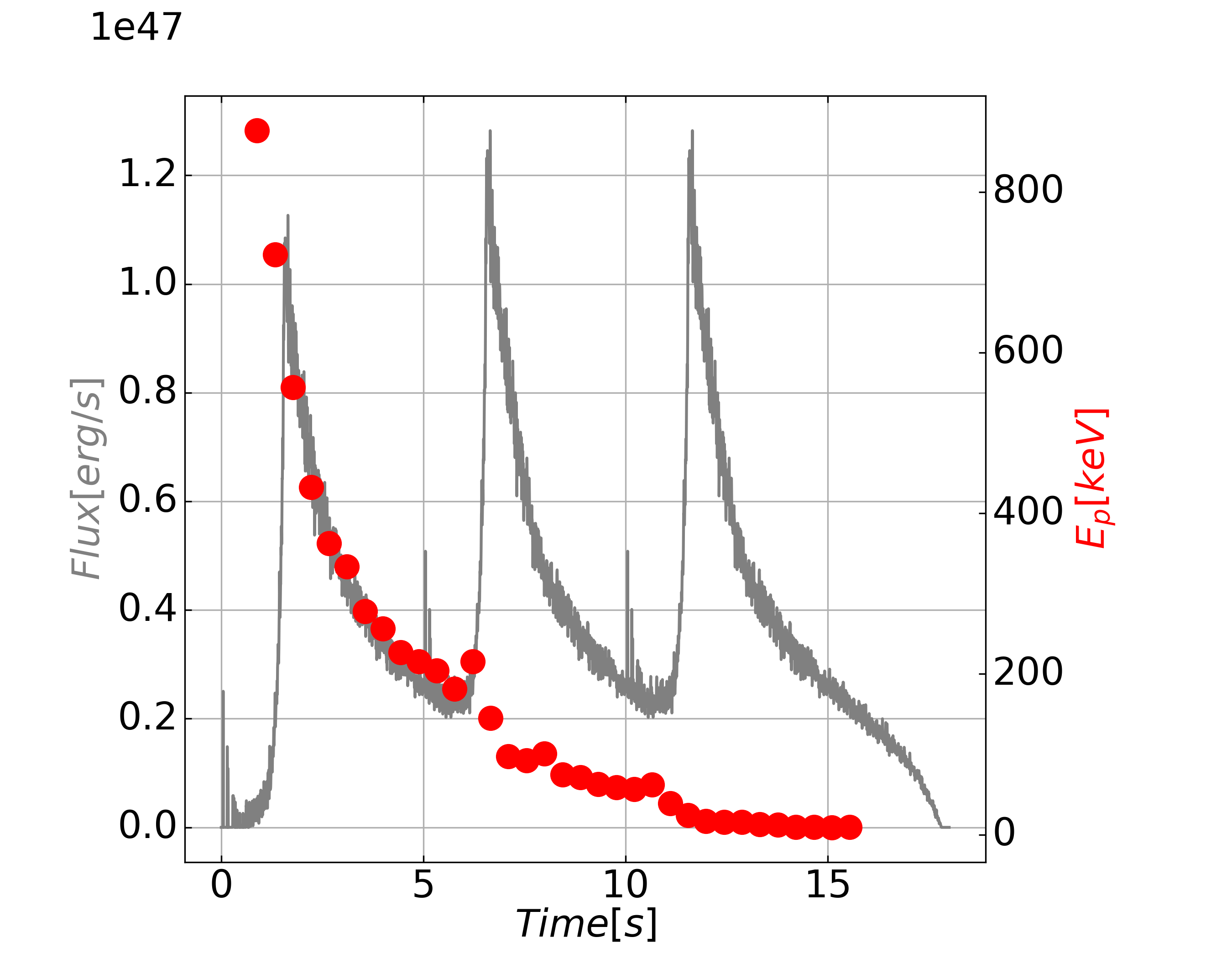}
    \label{fig:zero peak}
    }
    \subfigure[]{
    \includegraphics[width=0.3\textwidth]{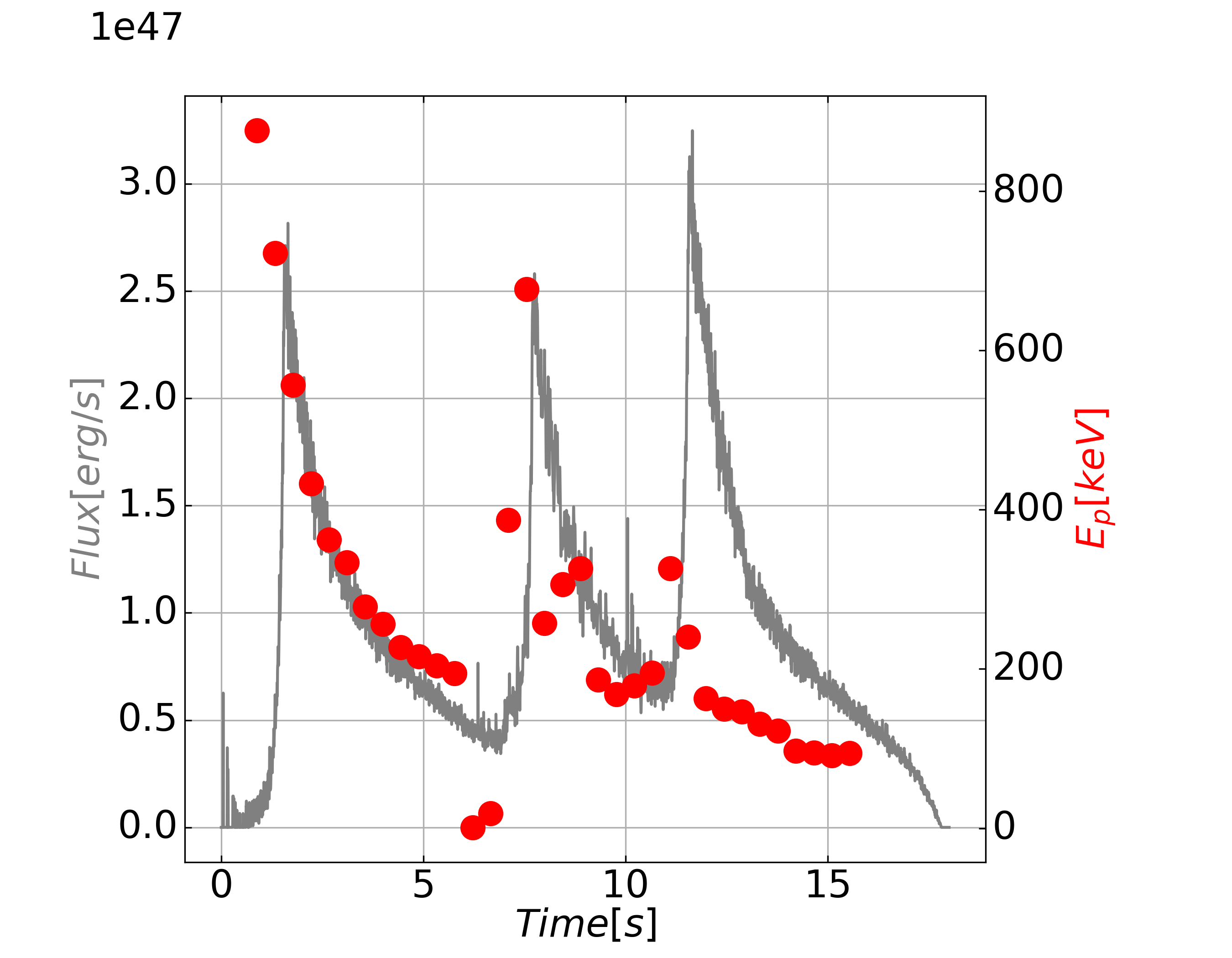}
    \label{fig:dud}
    }
    \subfigure[]{
    \includegraphics[width=0.3\textwidth]{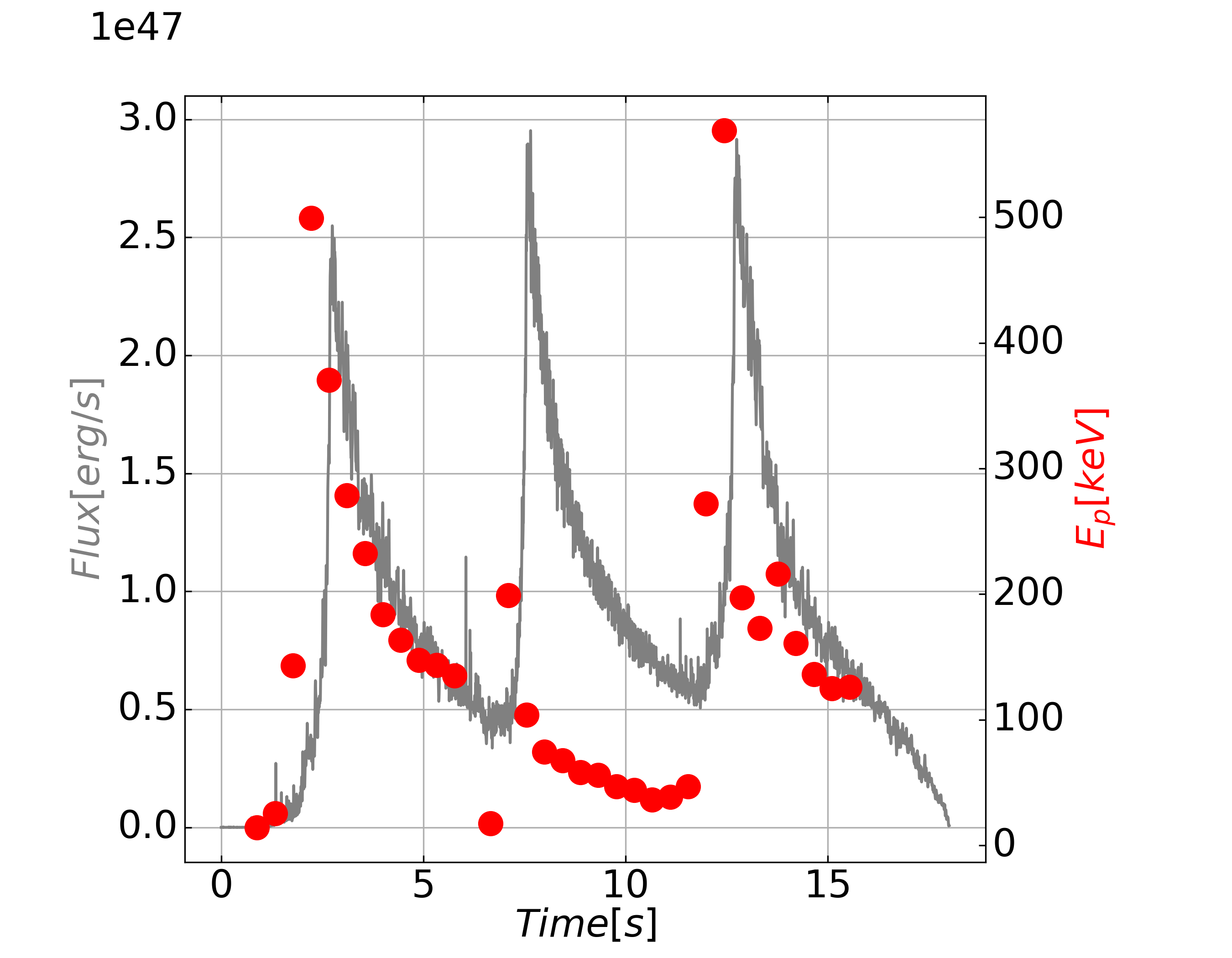}
    \label{fig:udu}
    }
    \caption{
        Simulation for different \Ep evolution patterns by taking results in \cref{fig:hard-to-soft E_peak evolution,fig:tracking E_peak evolution} as templates: (a) overall intensity-tracking; (b) first hard-to-soft then intensity-tracking; (c) first intensity-tracking then hard-to-soft; (d) overall hard-to-soft; (e) and (f): two patterns appear alternately.
        The grey lines represent the light curves and the red spots represent the \Ep value.
    }
    \label{fig:different patterns of E_peak}
\end{figure}

For one ICMART event, we find that the spectral peak \Ep evolution is always a hard-to-soft pattern, namely \Ep decreases throughout the whole event, even during the rising phase of the slow component, as shown in \cref{fig:hard-to-soft E_peak evolution}.
This is mainly because $E_\text{p} \propto B''_\text{e}$ and the magnetic field decreases with radius (and hence, time) as the shell expands in space.
Our result is consistent with previous works \citep{Uhm2016,LucasUhm2018}, in which emitter are from a signal large emission region with decaying magnetic field.

However, it is interesting to note that a GRB light curve is usually composed of multiple ICMART events that are fundamentally driven by the erratic GRB central engine activity. In this case, one broad component in real data is not necessarily generated by only one ICMART event.
In \cref{fig:tracking E_peak evolution}, we show an example light curve, which is generated with the superposition of three independent ICMART events.
The physical parameters for each ICMART event are listed in \cref{table:factors of tracking E_peak}.
Basically, by adjusting the parameters, we make the flux of the three events increase in turn over time.
If the absolute value of \Ep is positively correlated with flux, which seems to be supported by GRB observations (e.g., Amati/Yonetoku relations \citep{Amati2002,Yonetoku2004}), the \Ep evolution related to the overall light curve could be disguised as the intense-tracking pattern, namely \Ep increases during the rising phase of the broad pulse and decreases during the falling phase.

If our interpretation is correct, this may explain why both types of \Ep behavior are observed in individual GRB pulses.
Furthermore, for one particular GRB whose light curve consists of multiple broad pulses, the \Ep evolution for each broad pulse could be randomly hard-to-soft pattern or intense-tracking pattern.
This could naturally explain why mixed \Ep-evolution patterns can coexist in the same burst, with a variety of combined patterns \citep{Lu2012}.
For better illustration, we use the simulation results in \cref{fig:E_peak evolution of single peak in light curve} as templates to simulate the \Ep evolution and show that GRBs produced by multiple ICMART events could
\begin{enumerate}
    \item have all pulses (including the first one) showings intensity-tracking behavior (see \cref{fig:three peak});
    \item have the first pulse showing a clear hard-to-soft evolution, while the rest of the pulses showing tracking behavior (see \cref{fig:two peak});
    \item have the first pulse showing a nice tracking behavior, but the later pulses showing a clear hard to soft evolution (see \cref{fig:one peak});
    \item have a overall hard to soft evolution (see \cref{fig:zero peak});
    \item have two patterns appear alternately (see \cref{fig:udu,fig:dud}).
\end{enumerate}
It is noticing that all these patterns have already been observed in the current GRB sample \citep{Lu2012}.

\section{Conclusion and discussion}
\label{sec:conclusion}

The prompt emission of GRBs is still a mystery.
The main uncertainty is the composition of the outflow, Poynting flux dominated or matter-dominated, which essentially determines the energy dissipation mechanism, particle acceleration mechanism, and radiation mechanism.
Putting together all the observational evidence, including the properties of the light curve, the nature of the spectrum and the coordination between the spectrum and the light curve (e.g. the correlation between \Ep and flux), may help to solve the problem.

In this work, we have developed a numerical code to simulate the prompt emission light curves, time resolved spectrum and \Ep evolution patterns for GRBs produced by the Internal-Collision-induced MAgnetic Reconnection and Turbulence (ICMART) model.
Our simulation results could be summarized as follows:
\begin{itemize}
    \item The ICMART model could produce highly variable light curves, which can be decomposed as the superposition of an underlying slow component and a more rapid fast component.
    Such result is consistent with the previous simulations \citep{Zhang2014} and the observation facts \citep{Gao2012}.
    \item The ICMART model could produce a Band shape spectrum, whose parameters (\Ep, $\alpha$, $\beta$) could distribute in the typical distribution from GRB observations, as long as the magnetic field and the electron acceleration process in the emission region are under appropriate conditions.
    \item For one ICMART event, the spectral peak \Ep evolution is always hard-to-soft pattern. But if one individual broad pulse in the GRB light curve is composed of multiple ICMART events, the \Ep evolution related to the overall light curve could be disguised as the intense-tracking pattern.
    Therefore, mixed \Ep-evolution patterns can coexist in the same burst, with a variety of combined patterns.
\end{itemize}

Our results show that the ICMART model can explain the main characteristics of GRB light curve and spectrum, making it a very competitive model to produce GRB prompt emission.
However, to interpret GRB observations in great detail, the ICMART model still faces some difficulties, which are worth pointing out.

In principle, the total energy and magnetic field strength carried by the shell of ICMART event can vary randomly in a wide range. However, in order to make GRBs from ICMART events, it is required that for all magnetic reconnection process, there is a preferred range for the magnetic field and the electron acceleration process in the emission region.
Detailed numerical simulations of magnetic reconnection and particle acceleration processes are needed to address whether the \Be and \gm range demanded by the model could be achieved.
One possibility is that the magnetic reconnection process is so intense that there is an upper limit on the strength of \Be (e.g. smaller than $10^4$ G).
It is worth noticing that for most GRBs, \Be can not reach this upper limit, but needs to be lower than 100 G and needs to decay very rapidly (down 1-2 orders of magnitude) during the radiation process, otherwise the lower energy index $\alpha$ for majority of GRBs would distribute around $-1.5$ (entering deep fast cooling regime), instead of around $-1$ as suggested by the observations.

It is possible that in some ICMART event, \Be is even lower than 10 G.
For these cases, \Ep of the spectrum would shift into the soft X-ray band.
This might be the physical origin for the phenomenon so called ``X-ray flashes (XRFs)", which are generally believed not to be a different population from GRBs, but rather the natural extension of GRBs to the softer, less luminous regime \citep{Sakamoto2008}.   %
Accumulated sample of more XRFs in the future with sky survey detectors in the X-ray band (e.g. Einstein Probe; \citep{Yuan2018}) would help to justify this hypothesis.

\begin{acknowledgments}
    We thank helpful discussion with Bing Zhang. This work is supported by the National Natural Science Foundation of China (NSFC) under Grant No. 12021003.
    
\end{acknowledgments}

\bibliography{paper.bib}

\end{document}